\documentclass[english,aps,prl,twocolumn]{revtex4}%
\usepackage[T1]{fontenc}
\usepackage[latin9]{inputenc}
\usepackage{amsmath}
\usepackage{amssymb}
\usepackage{graphicx}
\usepackage{amsfonts}
\usepackage{times}
\usepackage{babel}%
\makeatletter
\@ifundefined{textcolor}{}
{
\definecolor{BLACK}{gray}{0}
\definecolor{WHITE}{gray}{1}
\definecolor{RED}{rgb}{1,0,0}
\definecolor{GREEN}{rgb}{0,1,0}
\definecolor{BLUE}{rgb}{0,0,1}
\definecolor{CYAN}{cmyk}{1,0,0,0}
\definecolor{MAGENTA}{cmyk}{0,1,0,0}
\definecolor{YELLOW}{cmyk}{0,0,1,0}
}
\makeatother
\begin{document}
\title{Role of the sampling weight in evaluating classical time autocorrelation functions}
\author{Tom\'{a}\v{s} Zimmermann}
\email{tomas.zimmermann@epfl.ch}
\author{Ji\v{r}\'{\i} Van\'{\i}\v{c}ek}
\email{jiri.vanicek@epfl.ch}
\affiliation{Laboratory of Theoretical Physical Chemistry, Institut des Sciences et
Ing\'{e}nierie Chimiques, Ecole Polytechnique F\'{e}d\'{e}rale de Lausanne,
Lausanne, Switzerland}

\begin{abstract}
We analyze how the choice of the sampling weight affects the efficiency of the
Monte Carlo evaluation of classical time autocorrelation functions. Assuming
uncorrelated sampling or sampling with constant correlation length, we propose
a sampling weight for which the number of trajectories needed for convergence
is independent of the correlated quantity, dimensionality, dynamics, and
phase-space density. In contrast, it is shown that the computational cost of
the \textquotedblleft standard\textquotedblright\ intuitive algorithm which
samples directly from the phase-space density may scale exponentially with the
number of degrees of freedom. Yet, for the stationary Gaussian distribution of
harmonic systems and for the autocorrelation function of a linear function of
phase-space coordinates, the computational cost of this standard algorithm is
also independent of dimensionality.

\end{abstract}
\maketitle

\selectlanguage{english}

\selectlanguage{english}

\emph{Introduction: Time-correlation functions. }Many dynamical properties of
stationary systems as well as the response of such systems to weak
perturbations can be inferred from time autocorrelation functions
\citep{Nitzan2006,Berne1970}. Examples include the optical absorption line
shapes computed from the dipole time autocorrelation function, the diffusion
coefficient computed from the velocity time autocorrelation function, and
various relaxation properties \citep{Williams1972}. More general time
correlation functions are in fact the principal ingredients of semiclassical
\citep{Sun1999,Liu2011} and path-integral
\citep{cao:1994,Craig2004,miller_manolopoulos:2005,Habershon2008,witt:2009,Perez2009}
calculations of quantum dynamical properties. Trajectory-based methods for
computing time correlation functions, however, often become too expensive in
many-dimensional systems. Yet, dimensionality-independent algorithms have been
found for special correlation functions, such as classical \citep{Mollica2011}
and semiclassical \citep{Mollica2011a} fidelity \cite{gorin:2006}. Motivated
by the success in these special cases and by the importance of correlation
functions in many areas of physics, in this Letter we explore how these
functions can be computed more efficiently in general. In particular, we
propose a sampling weight for which the number of trajectories needed for
convergence of any classical normalized time autocorrelation function is
independent of dimensionality.

Quantum mechanically, the unnormalized time autocorrelation function
$C_{\mathrm{u}}^{\mathrm{QM}}\left(  t\right)  $ of a vector operator
$\mathbf{\hat{A}}$ may be written as
\begin{equation}
C_{\mathrm{u}}^{\mathrm{QM}}\left(  t\right)  =\operatorname{Tr}(\hat{\rho
}^{0}\mathbf{\hat{A}}^{0}\cdot\mathbf{\hat{A}}^{t}), \label{eq:corr_func_qm}%
\end{equation}
where $\hat{\rho}^{0}$ is the density operator of the state, $\mathbf{\hat{A}%
}^{0}$ is the operator evaluated at time $t=0$, $\mathbf{\hat{A}}^{t}%
=e^{i\hat{H}t/\hbar}\mathbf{\hat{A}}e^{-i\hat{H}t/\hbar}$ is the operator
$\mathbf{\hat{A}}$ evolved with Hamiltonian $\hat{H}$ for time $t$, and
subscript \textquotedblleft u\textquotedblright\ emphasizes that the
correlation function is not normalized. The classical analog $C_{\mathrm{u}%
}^{\mathrm{CL}}\left(  t\right)  $ of the quantum correlation function
(\ref{eq:corr_func_qm}) is
\begin{equation}
C_{\mathrm{u}}^{\mathrm{CL}}\left(  t\right)  =h^{-D}\int dx\rho^{0}\left(
x\right)  \mathbf{A}^{0}\left(  x\right)  \cdot\mathbf{A}^{t}\left(  x\right)
, \label{eq:corr_func_cl}%
\end{equation}
where $x:=\left(  q,p\right)  $ is the $2D$-dimensional phase-space
coordinate, $\rho^{0}\left(  x\right)  $ is the initial phase-space density,
$\mathbf{A}^{0}\left(  x\right)  $ is the classical observable $\mathbf{A}$
evaluated at time $t=0$, and $\mathbf{A}^{t}\left(  x\right)  =e^{-\hat{L}%
t}\mathbf{A}^{0}\left(  x\right)  $ is this function $\mathbf{A}$ evolved
classically for time $t$ with the Liouville operator $\hat{L}=\left\{
H,\cdot\right\}  $. Note that besides a 3-dimensional vector (such as the
molecular dipole $\boldsymbol{\mu}$), $\mathbf{A}$ can also be a scalar ($A$)
or a higher-dimensional phase-space vector. To make the connection between
classical and quantum mechanical expressions explicit, the phase-space volume
is measured in units of $h^{D}$. Since our focus is on classical correlation
functions, superscript $\mathrm{CL}$ will be omitted from this point forward.

The shape of the autocorrelation function is often more interesting than its
overall magnitude \citep{Mcquarrie1976}, and hence one often computes the time
autocorrelation $C(t)$ which is normalized with respect to its initial value:%

\begin{equation}
C\left(  t\right)  =\frac{C_{\mathrm{u}}\left(  t\right)  }{C_{\mathrm{u}%
}\left(  0\right)  }. \label{eq:corr_func_cl_norm}%
\end{equation}

\emph{Algorithms. }Most common methods for evaluating
Eqs.~(\ref{eq:corr_func_cl}) and (\ref{eq:corr_func_cl_norm}) in
many-dimensional cases are based on classical trajectories. Two general
approaches are currently used \citep{Tuckerman2010}: 1) the direct approach in
which initial conditions for many trajectories are sampled from the stationary
distribution $\rho$ and the trajectories are subsequently evolved
simultaneously in time; and 2) the single-trajectory approach in which only
one trajectory is evolved in time and the desired autocorrelation function is
computed as an average of many correlation functions computed using the same
trajectory but initiated at different times. The direct approach is more
general and does not require the ergodicity of the time evolution, whereas the
single trajectory approach is generally simpler as it avoids explicit sampling
of $\rho$. Here we explore modifications of the direct approach using
generalized sampling weights.

We start by expressing the correlation function \eqref{eq:corr_func_cl} in
terms of trajectories,
\begin{equation}
C_{\text{u}}\left(  t\right)  =h^{-D}\int dx^{0}\rho\left(  x^{0}\right)
\mathbf{A}\left(  x^{0}\right)  \cdot\mathbf{A}\left(  x^{-t}\right)  ,
\label{eq:corr_func_trajs}%
\end{equation}
where $x^{t}:=\Phi^{t}(x^{0})$ is the phase-space coordinate at time $t$ of a
trajectory of the Hamiltonian flow $\Phi^{t}$ with initial condition $x^{0}$.
We further rewrite Eq.~\eqref{eq:corr_func_trajs} in a form suitable for Monte
Carlo evaluation, i.e., as an average
\begin{equation}
\left\langle E(x^{0},t)\right\rangle _{W}:=\frac{\int dx^{0}E(x^{0}%
,t)W(x^{0})}{\int dx^{0}W(x^{0})}, \label{eq:Monte_Carlo_average}%
\end{equation}
where the positive definite function $W$ is the sampling weight and $E$ is the
estimator. In the Monte Carlo method, average \eqref{eq:Monte_Carlo_average}
is evaluated numerically as an average
\begin{equation}
E_{W}(N,t)=\frac{1}{N}\sum_{j=1}^{N}E\left(  x_{j}^{0},t\right)
\label{eq:Monte_Carlo_sum}%
\end{equation}
over $N$ trajectories whose initial conditions $x_{j}^{0}$ are sampled from
the weight $W$.

The convergence rate of the sum \eqref{eq:Monte_Carlo_sum} usually depends on
$W$. Among many possible weights $W$, the form of Eq.\thinspace
\eqref{eq:corr_func_trajs} immediately suggests the following three:
$W(x)=\rho(x),$ $W(x)=\rho(x)\left\vert \mathbf{A}(x)\right\vert $, and
$W(x)=\rho(x)\mathbf{A}(x)^{2}$. These three weights lead to three different
algorithms, which may be generally written as
\begin{equation}
C_{\text{u,}W}\left(  t\right)  =I_{W}\,\left\langle E_{W}\left(
x^{0},t\right)  \right\rangle _{W}, \label{eq: Monte_Carlo_with_pref}%
\end{equation}
where $I_{W}:=h^{-D}\int W(x)dx$ denotes the norm of $W$ and the estimators
are
\begin{align}
E_{\rho}\left(  x^{0},t\right)   &  =\mathbf{A}\left(  x^{0}\right)
\cdot\mathbf{A}\left(  x^{-t}\right)  ,\label{eq:estimator_rho}\\
E_{\rho\left\vert \mathbf{A}\right\vert }\left(  x^{0},t\right)   &
=\frac{\mathbf{A}\left(  x^{0}\right)  \cdot\mathbf{A}\left(  x^{-t}\right)
}{\left\vert \mathbf{A}\left(  x^{0}\right)  \right\vert }%
,\label{eq:estimator_rhoA}\\
E_{\rho\mathbf{A}^{2}}\left(  x^{0},t\right)   &  =\frac{\mathbf{A}\left(
x^{0}\right)  \cdot\mathbf{A}\left(  x^{-t}\right)  }{|\mathbf{A}\left(
x^{0}\right)  |^{2}}. \label{eq:estimator_rhoA2}%
\end{align}
Substitution of Eq.~\eqref{eq: Monte_Carlo_with_pref} into the definition
(\ref{eq:corr_func_cl_norm}) yields a Monte Carlo prescription for the
normalized correlation function:
\begin{equation}
C_{W}\left(  t\right)  =\frac{\left\langle E_{W}\left(  x^{0},t\right)
\right\rangle _{W}}{\left\langle E_{W}\left(  x^{0},0\right)  \right\rangle
_{W}}. \label{eq:Monte_Carlo_complete}%
\end{equation}
Note that since $E_{\rho\mathbf{A}^{2}}\left(  x^{0},0\right)  =1$, no
normalization is needed for the $\rho\mathbf{A}^{2}$ algorithm. The two
averages in Eq.~\eqref{eq:Monte_Carlo_complete} may be evaluated either with
two independent Monte Carlo simulations or during a single Monte Carlo
simulation. Here we consider only the latter possibility, as it is
computationally faster and normalizes both $C_{\rho}\left(  0\right)  $ and
$C_{\rho\left\vert \mathbf{A}\right\vert }\left(  0\right)  $ exactly.

\emph{Statistical errors.} The three algorithms differ by the sampling weight
$W$ used and consequently also by the estimator $E_{W}$. The computational
cost of all three algorithms is $O\left(  c\frac{t}{\Delta t}N\right)  $,
where $N$ is the number of trajectories, $\Delta t$ the time step used, and
$c$ the combined cost of a single evaluation of the force (needed for the
dynamics) and of the estimator $E_{W}$. Usually, the cost of evaluating the
estimator is or can be made negligible to that of evaluating the force.
Therefore the costs of the algorithms differ mainly in the number $N$ of
trajectories needed to achieve a desired precision (i.e., discretization
error) $\sigma_{\text{discr}}$.

Alternatively, the algorithms can be compared by evaluating the discretization
errors $\sigma_{\text{discr,}W}$ resulting from a given number $N$ of
trajectories. For an unbiased estimator, the discretization error
$\sigma_{\text{discr}}$ is equal to the statistical error $\sigma_{W}$, where
$\sigma_{W}(N,t)^{2}={\overline{C_{W}(N,t)^{2}}-\overline{C_{W}(N,t)}^{2}}$
and the overline denotes an average over an infinite number of simulations
with different sets of $N$ trajectories. Assuming for now that the $N$
trajectories are uncorrelated, one can show that the error of the unnormalized
$C_{\text{u}}(t)$ satisfies%
\begin{equation}
\sigma_{\text{u,}W}(N,t)^{2}=\frac{I_{W}^{2}}{N}\left[  \left\langle
E_{W}\left(  x^{0},t\right)  ^{2}\right\rangle _{W}-\left\langle E_{W}\left(
x^{0},t\right)  \right\rangle _{W}^{2}\right]  . \label{eq:sigma_theor}%
\end{equation}
For $W=\rho\mathbf{A}^{2}$, the error of normalized $C(t)$ satisfies an
analogous relation obtained by removing factors of $I_{W}$ from Eq.
(\ref{eq:sigma_theor}). Statistical errors of algorithms with weights $\rho$
and $\rho\left\vert \mathbf{A}\right\vert $, which must be normalized
according to Eq.~\eqref{eq:Monte_Carlo_complete}, are found from the formula
for the statistical error of a ratio of random variables:
\begin{equation}
\left(  \frac{\sigma_{S/T}}{\overline{S/T}}\right)  ^{2}=\left(  \frac
{\sigma_{S}}{\bar{S}}\right)  ^{2}+\left(  \frac{\sigma_{T}}{\bar{T}}\right)
^{2}-2\frac{\overline{ST}-\bar{S}\bar{T}}{\bar{S}\bar{T}}.
\label{error_of_ratio}%
\end{equation}
In our case, $S=C_{\text{u,}W}\left(  N,t\right)  $ and $T=C_{\text{u,}%
W}\left(  N,0\right)  $. Realizing that $\overline{C_{\text{u,}W}%
(N,t)}=C_{\text{u}}(t)$ we obtain the following general expression for the
statistical errors of the three algorithms:
\begin{equation}
\sigma_{W}\left(  N,t\right)  ^{2}=\frac{1}{Nd_{W}}\left[  a_{W}C\left(
t\right)  ^{2}-2b_{W}C\left(  t\right)  +c_{W}\right]  , \label{eq:error}%
\end{equation}
where $a_{\rho}=\langle\left\vert \mathbf{A}^{0}\right\vert ^{4}\rho
/W\rangle_{\rho}$, $b_{\rho}=\langle\left\vert \mathbf{A}^{0}\right\vert
^{2}\left(  \mathbf{A}^{0}\cdot\mathbf{A}^{t}\right)  \rho/W\rangle_{\rho}$,
$c_{\rho}=\langle(\mathbf{A}^{0}\cdot\mathbf{A}^{t})^{2}\rho/W\rangle_{\rho}$,
$d_{\rho}=\langle\left\vert \mathbf{A}^{0}\right\vert ^{2}\rho/W\rangle_{\rho
}^{2}$, and an abbreviated notation $\mathbf{A}^{t}:=\mathbf{A}(x^{-t})$ was
used. The special cases are obtained by replacing $W$ with $\rho$,\textbf{
}$\rho\left\vert \mathbf{A}\right\vert $, or $\rho\mathbf{A}^{2}$ in these expressions.

For $W=\rho\mathbf{A}^{2}$, the coefficients can be rearranged as
$a_{\rho\mathbf{A}^{2}}=-d_{\rho\mathbf{A}^{2}}$, $b_{\rho\mathbf{A}^{2}}=0$,
$c_{\rho\mathbf{A}^{2}}=\left\langle (\mathbf{A}^{0}\cdot\mathbf{A}^{t}%
)^{2}/|\mathbf{A}^{0}|^{2}\right\rangle _{\rho}$, and $d_{\rho\mathbf{A}^{2}%
}=\left\langle |\mathbf{A}^{0}|^{2}\right\rangle _{\rho}$. Using the
Cauchy-Schwarz inequality $(\mathbf{A}^{0}\cdot\mathbf{A}^{t})^{2}%
\leq|\mathbf{A}^{0}|^{2}|\mathbf{A}^{t}|^{2}$ in the expression for
$c_{\rho\mathbf{A}^{2}}$ and the fact that for stationary distributions
$\left\langle |\mathbf{A}^{0}|^{2}\right\rangle _{W}=\left\langle
|\mathbf{A}^{t}|^{2}\right\rangle _{W}$, we find that $c_{\rho\mathbf{A}^{2}%
}\leq\langle\mathbf{A}\left(  x^{-t}\right)  ^{2}\rangle_{\rho}=d_{\rho
\mathbf{A}^{2}}$ and realize that for the weight $\rho\mathbf{A}^{2}$ the
upper bound for the statistical error depends only on $N$ and the value of the
autocorrelation function $C\left(  t\right)  $:
\begin{equation}
\sigma_{\rho\mathbf{A}^{2}}^{2}\left(  N,t\right)  \leq\frac{1}{N}[1-C\left(
t\right)  ^{2}]. \label{eq:sigma_rhoA2}%
\end{equation}
In particular, the error does not explicitly depend on the dimensionality $D$
of the system, chaoticity of its dynamics, the nature of the observable
$\mathbf{A}$, or time $t$. This remarkable fact is the main thesis of this paper.

\textit{Special cases.} One cannot make a similar general statement about
either of the algorithms using weight $\rho$\textbf{ }or\textbf{ }%
$\rho\left\vert \mathbf{A}\right\vert .$ We therefore turn to two special
cases permitting analytical evaluation of the statistical errors. Both
examples involve a many-dimensional harmonic oscillator (HO) $H=(1/2)(p^{2}%
/m+kq^{2})$ and its stationary Gaussian distribution
\begin{equation}
\rho(x)=[2\tanh(u/2)]^{D}\exp[-\tanh(u/2)(q^{2}/a^{2}+p^{2}a^{2}/\hbar^{2})],
\label{Wig_Boltz}%
\end{equation}
given by the Wigner transform of the Boltzmann density operator. Above,
$u:=\beta\hbar\omega$, $\omega^{2}=k/m$, $a^{2}=\hbar/(m\omega)$. [Note that
the ground state density and the classical Boltzmann distribution can be
obtained as the limits of Eq. (\ref{Wig_Boltz}) for $\beta\rightarrow\infty$
and $\beta\rightarrow0$, respectively.] The two examples differ in the choice
of the observable $\mathbf{A}$.

\textit{Exponential growth of }$\sigma$ \textit{with }$D$. First consider $A$
to be the product of coordinates: $A=q_{1}q_{2}\cdots q_{D}$. The statistical
error for $W=\rho A^{2}$ is described by Eq.\ \eqref{eq:sigma_rhoA2} in full
generality and thus is independent of $D$. On the other hand, straightforward
but somewhat tedious calculations using Eq. (\ref{eq:error}) show that
statistical errors for both weights $\rho$ and $\rho\left\vert A\right\vert $
grow exponentially with the number of dimensions $D$:
\begin{align}
\sigma_{\rho}\left(  N,t\right)  ^{2}  &  =\frac{1}{N}\left\{  \left[
1+2\sqrt[D]{C\left(  t\right)  ^{2}}\right]  ^{D}-3^{D}C\left(  t\right)
^{2}\right\}  ,\label{eq:product_sigma_rho}\\
\sigma_{\rho\left\vert A\right\vert }\left(  N,t\right)  ^{2}  &  =\frac{1}%
{N}\left(  \frac{2}{\pi}\right)  ^{D}\left\{  \left[  1+\sqrt[D]{C\left(
t\right)  ^{2}}\right]  ^{D}-2^{D}C\left(  t\right)  ^{2}\right\}  .
\label{eq:product_sigma_rhoA}%
\end{align}
The fact that for $W=\rho$ and $\rho\left\vert A\right\vert $ there exist
observables for which the error grows exponentially with $D$ is our second
main result. Similar behavior of $\sigma$ is expected for any multiplicatively
separable function $A$ of phase-space coordinates, such as the Gaussian
$A=\exp(-q^{2}/a^{2})$.

\textit{Independence of }$D$. Yet, the situation is not always so bleak.
Consider the correlated function $A=\mu^{\prime}\cdot q$ to be a linear
function of coordinates $q$ ($\mu^{\prime}$ is a $D$-dimensional vector). In
this important special case, all three sampling methods have statistical
errors independent of dimensionality:%
\begin{align}
\sigma_{\rho\text{ or }\rho A^{2}}\left(  N,t\right)  ^{2}  &  =\frac{1}%
{N}[1-C\left(  t\right)  ^{2}],\label{eq:sigma_linear_dipole}\\
\sigma_{\rho\left\vert A\right\vert }\left(  N,t\right)  ^{2}  &  =\frac
{2}{\pi N}[1-C\left(  t\right)  ^{2}]. \label{eq:sigma_linear_dipole_rhoA}%
\end{align}
The proof of Eq. (\ref{eq:sigma_linear_dipole_rhoA}) for weight $\rho
\left\vert A\right\vert $ is somewhat involved and was done only for the case
$\mu_{1}=\cdots=\mu_{D}$. On the other hand, Eq. (\ref{eq:sigma_linear_dipole}%
) remains valid even for HOs with different frequencies in different
dimensions. Note that the statistical error is slightly lower for
$W=\rho\left\vert A\right\vert $ than for $W=\rho$ or $\rho A^{2}$.

\textit{Sampling methods and correlation length.} Before presenting numerical
examples, let us briefly discuss the sampling methods. In many dimensions,
sampling from a general weight $W$ is often performed with the Metropolis
method \citep{Metropolis1953,Hastings1970,Chib1995}. Two variants are used
here: The \textquotedblleft original\textquotedblright\ Metropolis method
proposes the new point $x_{\text{new}}$ using a random walk step from the last
accepted point $x_{\text{old}}$; $x_{\text{new}}$ is accepted with probability
$p_{\mathrm{acc}}=\min[W(x_{\mathrm{new}})/W(x_{\mathrm{old}}),1]$. If
$x_{\text{new}}$ is rejected, the last accepted point $x_{\text{old}}$ is
duplicated. In the \textquotedblleft product\textquotedblright\ Metropolis
method, $W$ is factorized as $W=YZ$, where $Y$ can be sampled
\textquotedblleft directly\textquotedblright\ to propose a new point
$x_{\text{new}}$ which is subsequently accepted with probability
$p_{\mathrm{acc}}=\min[Z(x_{\mathrm{new}})/Z(x_{\mathrm{old}}),1]$.

Unfortunately, except for a few distributions $W$ (such as the uniform or
normal distributions, which may be sampled \textquotedblleft
directly\textquotedblright), points generated by Metropolis methods are
correlated, leading to a correlation length $N_{\mathrm{corr}}>1$ between
samples. This increases the statistical error for a given number of samples
$N$. As a consequence, in all of our analytical expressions, $N$ should be
replaced by $N/N_{\text{corr}}$, which can affect (slightly) the dependence of
$\sigma$ on $D$. An important factor increasing $N_{\mathrm{corr}}$ is the
rejection of proposed moves, which results in exactly identical samples. In a
properly designed code, however, these repeated samples do not increase the
computational cost; they are accounted for by increasing the statistical
weight of the original (not yet duplicated) sample. Thus, strictly speaking,
the efficiency of a sampling algorithm depends on the number $N_{\mathrm{uniq}%
}$ of unique trajectories needed for convergence rather than on the total
number $N$ of trajectories. While we took $N_{\text{corr}}$ into account in
the numerical calculations, a detailed analysis of $N_{\text{corr}}$, which
can both increase (slowly) or decrease (slowly) with $D$, is beyond the scope
of this paper.

\textit{Numerical results.} We first confirmed our analytical results for HOs
numerically using $k=m=\hbar=\beta=1.$ Numerical statistical errors were
estimated by averaging these errors over $100$ independent simulations, each
with the same number of unique trajectories $N_{\mathrm{uniq}}=5\times10^{5}$.
In order to compare with the analytical results, the effect of correlation was
removed by converting the numerical statistical error $\sigma$ to an error per
trajectory $\sigma_{1}:=(N/N_{\text{corr}})^{1/2}\sigma$. The correlation
lengths $N_{\mathrm{corr}}$ were estimated using the method of block averages \citep{Flyvbjerg1989}.

\begin{figure}[ptb]
\includegraphics[width=\columnwidth]{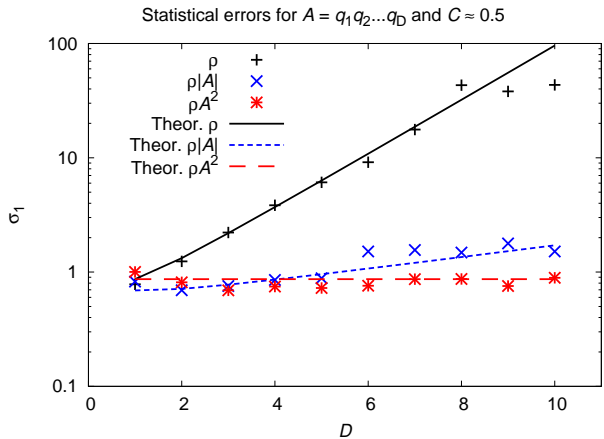}\caption{Expected
statistical error per trajectory of the autocorrelation function $C\left(
t\right)  $ of the function $A=q_{1}q_{2}\cdots q_{D}$ in a many-dimensional
harmonic oscillator. The statistical error is independent of dimensionality
for the algorithm with weight $W=\rho A^{2}$ and grows exponentially with $D$
for the other two weights. Time $t$ was chosen separately for each $D$ so that
$C(t)\approx0.5$.}%
\label{fig:Product-dipole-moment}%
\end{figure}

Figure\ \ref{fig:Product-dipole-moment} shows that for $A=q_{1}q_{2}\cdots
q_{D}$, the error $\sigma_{1}$ grows exponentially with $D$ for both weights
$\rho$ and $\rho|A|$ while it is independent of $D$ for $W=\rho A^{2}$.
Moreover, numerical results agree with the analytical predictions
(\ref{eq:sigma_rhoA2}), (\ref{eq:product_sigma_rho}), and
(\ref{eq:product_sigma_rhoA}). The original Metropolis method was used since
the acceptance rate of the product Metropolis method was prohibitively low for
high $D$. The step size of the random walk was the same for all three weights
but varied weakly with $D$ for the sake of a reasonable acceptance rate. [Note
that in our calculations $\sigma_{\rho A^{2}}=(N_{\text{corr}}/N)^{1/2}%
\sigma_{1,\rho A^{2}}$ itself grew slightly with $D$ due to a slow growth of
the correlation length $N_{\mathrm{corr}}$ with $D$. For $W=\rho$,
$N_{\mathrm{corr}}$ decreased slightly with $D$ and for $W=\rho\left\vert
A\right\vert $ it stayed approximately constant, but these effects did not
cancel the overall exponential growth of the error. Even though
$N_{\text{corr}}$ can be varied to some extent by modifying the step size of
the random walk, this was not explored in detail here.]

\begin{figure}[ptb]
\includegraphics[width=\columnwidth]{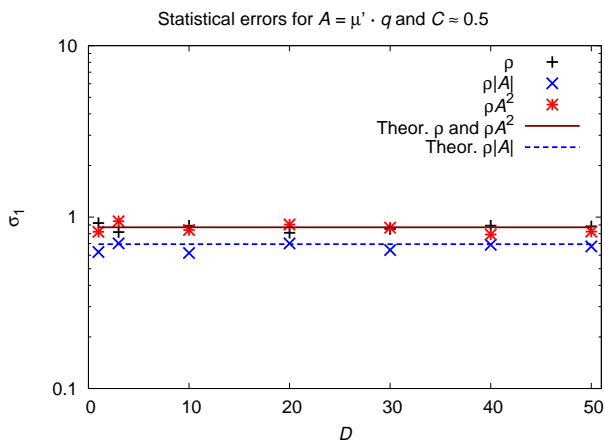}\caption{Expected
statistical error per trajectory of the autocorrelation function $C(t)$ of the
linear operator $A=\mu^{\prime}\cdot q$ in a many-dimensional harmonic
oscillator. The statistical error is independent of dimensionality for all
three sampling weights studied. Time $t$ was chosen separately for each $D$ so
that $C(t)\approx0.5$.}%
\label{fig:Linear-dipole-moment}%
\end{figure}

Figure\ \ref{fig:Linear-dipole-moment} compares the analytical predictions
with numerically computed errors for $A=\mu^{\prime}\cdot q$, where
$\mu^{\prime}$ is a $D$-dimensional vector with all entries equal to $1.$ Such
$A$ can be interpreted as a linear approximation to the electric dipole of a
nonpolar molecule. Figure \ref{fig:Linear-dipole-moment} confirms that the
statistical error $\sigma_{1}$ is independent of $D$ for all three algorithms.
Initial conditions were sampled using the product Metropolis algorithm with
$W=YZ$ and $Y=\rho$ in all cases. Function $Z$ used in the acceptance
criterion was equal to $1$, $\left\vert A\right\vert $, and $A^{2}$, for
$W=\rho$, $\rho\left\vert A\right\vert $, and $\rho A^{2}$, respectively.
Therefore, for $W=\rho$, $N_{\mathrm{corr}}=1$ and $N=N_{\mathrm{uniq}}$,
while for $W=\rho\left\vert A\right\vert $ and $\rho A^{2}$, $N_{\mathrm{corr}%
}>1$ and $N>N_{\mathrm{uniq}}$.

Finally, we used the three algorithms to calculate the vibrational spectrum of
a $48$-dimensional harmonic model of the ground electronic state of azulene
computed at the CASSCF(4,6)/6-31G{*} level of theory. Observable $\mathbf{A}$
was a linear approximation of the dipole moment of azulene, $\mathbf{A}%
=\boldsymbol{\mu}=\boldsymbol{\mu}_{0}+\boldsymbol{\mu}^{\prime}\cdot q$,
where $\boldsymbol{\mu}_{0}:=\boldsymbol{\mu}(0)$ is the equilibrium dipole
moment (a $3$-dimensional vector) and $\boldsymbol{\mu}^{\prime}$ the $3\times
D$ matrix of derivatives of the dipole moment at $q=0$. Sampling was performed
the same way as in the previous example. The dipole autocorrelation function
$C(t)$ was computed intentionally only up to time $t_{\text{tot}}=1.45%
\operatorname{ps}%
$, which is the minimum time needed to resolve all vibrational peaks, and with
only $N_{\text{uniq}}=10^{4}$ unique trajectories, for which $C(t)$ starts to
converge. Prior to computing the spectrum via a Fourier transform, $C(t)$ was
damped by a multiplication with the function $\cos(\pi t/2t_{\text{tot}})^{2}%
$. After the transform, $\mathcal{F}\left[  C\left(  t\right)  \right]
\left(  \omega\right)$ was multiplied by the factor $2\omega\tanh\left(
\frac{\beta\hbar\omega}{2}\right)$, which includes the standard
\textquotedblleft quantum correction\textquotedblright\ \cite{Berne1970} for
the lack of detailed balance in the classical $C\left(  t\right)$. While
this correction is \emph{not} exact even for HOs if $\rho$ is the classical
Boltzmann density, it becomes exact for harmonic systems if $\rho$ is the
Wigner Boltzmann density (\ref{Wig_Boltz}). Figure 3, showing the
high-frequency region of the spectrum containing the C-H bond stretches,
confirms that all three algorithms converge to the same result (agreeing,
within the resolution, with the exact spectrum). Moreover, even in this
slightly more general case than the one considered in Fig. 2, the statistical
errors associated with all three sampling weights stayed approximately
independent of $D.$ (Systems with $D<48$ were generated by progressively
cutting off the lowest frequency normal modes of azulene.) 

\begin{figure}[ptb]
\includegraphics[width=\columnwidth]{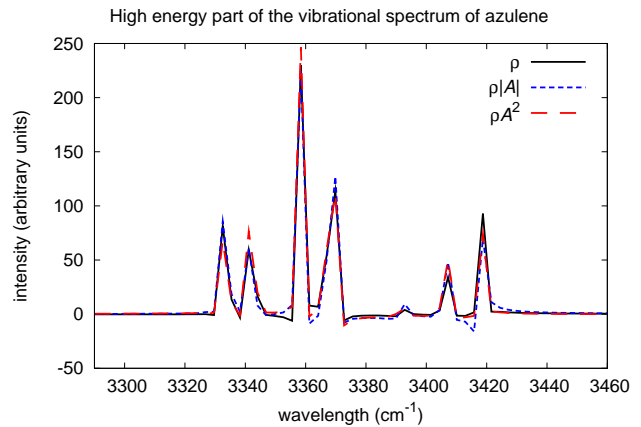}\caption{The
high frequency part of the vibrational spectrum of a harmonic model of azulene
computed via the Fourier transform of the dipole time autocorrelation
function.}%
\label{fig:spectrum}%
\end{figure}

\textit{Conclusions.} We have demonstrated the existence of a sampling weight
for which the number of trajectories needed for convergence of the normalized
time autocorrelation function of any phase-space function $\mathbf{A}$ is
independent of the dimensionality and the underlying dynamics of the system.
This sampling weight is $W=\rho\mathbf{A}^{2}$, which may not be surprising at
time $t=0$, when this $W$ represents the ideal importance sampling weight with
all trajectories contributing unity to the sum \eqref{eq:Monte_Carlo_sum}.
Here we have shown that this sampling weight retains its favorable properties
also for $t>0$ by proving that $\sigma_{\rho\mathbf{A}^{2}}$ depends
explicitly only on $C\left(  t\right)  $ itself, and not on other parameters
of the system.

While best suited for normalized autocorrelation functions, weight
$\rho\mathbf{A}^{2}$ can also accelerate calculations of unnormalized
autocorrelation functions $C_{\mathrm{u}}\left(  t\right)  $ via the relation
$C_{\text{u}}(t)=C_{\text{u}}(0)C(t)$. In the latter case, weight
$\rho\mathbf{A}^{2}$ is retained for the dynamical calculation of $C(t)$,
which is usually the most time-consuming task by far. The initial norm
$C_{\mathrm{u}}\left(  0\right)  $ must be computed separately using a
normalized sampling weight such as $\rho$. Yet, one can afford many more
trajectories for computing $C_{\mathrm{u}}\left(  0\right)  $ since this
calculation does not require any dynamics.

To conclude, we hope that the dimensionality-independent sampling weight will
find its use in other classical, semiclassical \citep{Sun1999,Liu2011}, and
even quantum mechanical trajectory-based applications, such as those using the
centroid \citep{cao:1994,Habershon2008,Perez2009,witt:2009} or ring-polymer
\citep{Craig2004,miller_manolopoulos:2005,Habershon2008,Perez2009,witt:2009}
molecular dynamics.

\textit{Acknowledgements. }This research was supported by the Swiss NSF with
grants No. 200021\_124936 and NCCR MUST, and by EPFL. We thank C. Mollica and
T. Prosen for discussions, and V. Sharma, D. Marcos Gonzalez, M. Wehrle, and
M. \v{S}ulc for assistance with numerical calculations.


\end{document}